
\magnification=1200\baselineskip=0.33truein.

\vskip2cm

\centerline{THE FINITE-TEMPERATURE FEYNMAN PROPAGATOR}

\centerline{IN OPERATOR FORM }

\vskip1cm

\centerline{H. Arthur Weldon}

\centerline{Department of Physics}

\centerline{West Virginia University}

\centerline{Morgantown, WV 26506-6315}

\vskip1cm
\centerline{September 1, 1995}

\vskip1cm

In momentum space the  Feynman propagator $D_{F}(k)$ at non-zero temperature is
defined by a simple dispersion relation with the
familiar property of being an even function of $k^{0}$ and analytic  for
Re$(k^{0})^{2}>0$.
The  coordinate space form of the propagator $D_{F}(x)$ is expressed directly
in
terms of matrix elements of the field operator and  requires a new type of
operator ordering.


\vfill\eject

\def\sqr#1#2{{\vcenter{\vbox{\hrule height.#2pt
			     \hbox{\vrule width.#2pt  height#1pt \kern#1pt
							   \vrule  width.#2pt}
        \hrule  height.#2pt}}}}
\def\square{\mathchoice\sqr74\sqr64\sqr{2.1}3\sqr{1.5}3}

\centerline{I. INTRODUCTION}

In zero-temperature field theory the Feynman propagator is the
vacuum expectation value of the time-ordered product of two field operators.
Therefore it is rather curious  that at non-zero temperature
 the  Feynman propagator and the time-ordered propagator
propagator  are  different.  At $T\neq 0$ the time-ordered propagator for a
real
scalar field $\phi$ is $$iD_{11}(x)=\sum_{N}\langle
N|T\bigl(\phi(x)\phi(0)\bigr)|N\rangle
\;{e^{-\beta E_{N}}\over Z}\eqno(1)$$
where  $|N\rangle$ are eigenstates of the total Hamiltonian and $\beta$ is the
inverse temperature. In momentum space it has the dispersion representation
[1,2]
$$D_{11}(k)=\int_{-\infty}^{\infty}d\omega\Bigl( {1+f(\omega)\over
k^{0}-\omega+i\eta} -{f(\omega)\over
k^{0}-\omega-i\eta}\Bigr)\;\rho(\omega,\vec{k})\eqno(2)$$ where
$f(\omega)=1/[e^{\beta\omega}-1]$ and  the spectral function is
$$\rho(k)={1\over 2\pi}\sum_{N}\int
d^{4}x\;e^{ik\cdot x} \langle N|\bigl[\phi(x),\phi(0)\bigr]|N\rangle
\;{e^{-\beta
E_{N}}\over Z}\eqno(3)$$
Because $\rho(k)$ is real, it is easy to separate the real and
imaginary parts of (2):
$$\eqalign{{\rm Re}\,D_{11}(k)&={\cal P}\int_{-\infty}^{\infty}d\omega\;
{\rho(\omega,\vec{k})\over k^{0}-\omega}\cr
{\rm Im}\,D_{11}(k)&=-\pi\coth({\beta k^{0}\over 2})\;\rho(k)\cr}\eqno(4)$$

To explain what is meant by the Feynman propagator at non-zero temperature it
is
helpful to  recall that real-time calculations require doubling the number of
degrees of freedom [2-9]. Associated with each physical field is an auxiliary
field. All propagators become $2\times 2$ matrices in the internal space. Thus
$D_{11}$ is one entry in the $2\times 2$ matrix. In momentum space it has the
 form [2-9]
 $$D_{ad}(k)=U_{ab}\left[\matrix{D_{F}(k) &0\cr
0& -D_{F}^{*}(k)}\right]_{bc}U_{cd}\eqno(5)$$
where $D_{F}(k)$ is the Feynman propagator. The simplicity of this matrix
structure indicates that the thermal Feynman propagator plays a central role in
thermal field theory.
To extract a representation for
$D_{F}(k)$, use $U_{12}=U_{21}=[\exp(\beta|k^{0}|)-1]^{-1/2}$ and
$U_{11}=U_{22}=\exp(\beta|k^{0}|/2)\,U_{12}$ to obtain
$$D_{11}(k)={D_{F}(k)\exp(\beta|k^{0}|)- D_{F}^{*}(k)\over
\exp(\beta|k^{0}|)-1}\eqno(6)$$
When the real and imaginary parts of this relation are compared to (4) the
result
is
$$\eqalign{{\rm Re}\,D_{F}(k)&={\cal P}\int_{-\infty}^{\infty}d\omega\;
{\rho(\omega,\vec{k})\over k^{0}-\omega}\cr
{\rm Im}\,D_{F}(k)&=-\pi\epsilon(k^{0})\;\rho(k)\cr}\eqno(7)$$
Consequently the thermal Feynman propagator satisfies the dispersion relation
$$D_{F}(k)=\int_{-\infty}^{\infty}d\omega{\rho(\omega,\vec{k})\over
k^{0}-\omega+i\eta\epsilon(k^{0})}\;\eqno(8)$$
Using $\rho(-k^{0},\vec{k})=-\rho(k^{0},\vec{k})$ this can also be written
$$D_{F}(k)=\int_{0}^{\infty}d\omega{2\omega\;\rho(\omega,\vec{k})\over
(k^{0})^{2}-\omega^{2}+i\eta}\;\eqno(9)$$
This dispersion relation is not new; it is discussed in [2] for example.
It shows that the thermal Feynman propagator is an even function of $k^{0}$
that is analytic in the region Re$(k^{0})^{2}>0$. The dispersion relation has
exactly the same appearance as at zero temperature because all dependence on
the
temperature is contained in the thermal spectral function. (In practice
it is more complicated than at zero temperature because   there are no
regions of $\omega$ where the  spectral function vanishes.)

 Since $D_{F}$ has rather simple
properties and plays a central role in the matrix (5), it seems worthwhile to
ask
how  $D_{F}$ can be expressed directly in terms of matrix elements of the field
operator $\phi(x)$. The answer to this question is given below in (10).
 The proof of (10) is that the Fourier
transform  produces the defining dispersion relation (8). The Appendix contains
a discussion of how to extract the operator form for the Feynman proper
self-energy $\Pi_{F}(x)$.

\vfill\eject

\centerline{II. OPERATOR FORM FOR $D_{F}(x)$}

In coordinate space the Feynman propagator is the thermal average of
a certain ordered product of two field operators. The ordering is, however,
more complicated than the usual time-ordering. The result is
$$\eqalign{iD_{F}(x)=\sum_{N}\Bigl[&\langle
N|\phi(x)\;\theta_{N}(t)\;\phi(0)|N\rangle\cr
 +&\langle
N|\phi(0)\;\theta_{N}(-t)\;\phi(x)|N\rangle\Bigr]
\;{e^{-\beta E_{N}}\over Z}\cr}\eqno(10)$$
The new ordering operation is defined by
$$\theta_{N}(t)=\theta(t)\theta(H-E_{N})
-\theta(-t)\theta(E_{N}-H)\eqno(11)$$
Having the Hamiltonian operator in the argument of the theta function is
 a compact way of  representing a projection operator. For example
$$\theta(H-E_{N})=\sum_{E_{A}>E_{N}}|A\rangle\langle A|\eqno(12)$$
Since the ordering depends only on the energy  $E_{N}$
and not on a particular state  $|N\rangle$, it
could be labelled $\theta_{E_{N}}(x^{0})$, but that seems cumbersome.
The time derivative of (11) is particularly simple:
$${d\over dt}\theta_{N}(t)=\delta(t)\eqno(13)$$
Note that (10) can be written in a variety of forms
by inserting a complete set of states (12) and regrouping the matrix elements
in various ways. In the zero-temperature limit (10) reduces to the usual
time-ordered product  as it should. In this limit the only state $|N\rangle$
that
contributes is the vacuum.
Since $\theta(H-E_{vac})=1$ and $\theta(E_{vac}-H)=0$,
it follows that $\theta_{N}(t)\to\theta(t)$ at zero temperature.

The remainder of the discussion will prove that the Fourier transform of (10)
yields the dispersion relation (8).
To do this it is simplest to transform only the time-dependence of (10) without
changing the space-dependence
$$D_{F}(k^{0},\vec{x})=\int_{-\infty}^{\infty}dt\;e^{ik^{0}t}\;
D_{F}(x)\eqno(14)$$
Examine the first term $\langle N|\phi(x)\theta_{N}(t)\phi(0)|N\rangle$ in
(10).
Using the time-dependence of the
Heisenberg field, $\phi(x)=\exp(iHt)\phi(\vec{x})\exp(-iHt)$, this can be
written
$$\langle N|\phi(x)\theta_{N}(t)\phi(0)|N\rangle =\langle
N|\phi(\vec{x})e^{i(E_{N}-H)t}\;\theta_{N}(t)\phi(0)|N\rangle$$ the Fourier
transform  is $$\int_{-\infty}^{\infty}dt\;e^{ik^{0}t}
\langle N|\phi(x)\theta_{N}(t)\phi(0)|N\rangle
=i\langle N|\phi(\vec{x})\;R\;\phi(0)|N\rangle$$
where
$$R\equiv{\theta(H-E_{N})\over k^{0}+E_{N}-H+i\eta}
+{\theta(E_{N}-H)\over k^{0}+E_{N}-H-i\eta}$$
As before, the appearance of the Hamiltonian operator can
be replaced by summing over a complete set of states.
One can write $R$ more compactly as
$$R=\int_{-\infty}^{\infty}d\omega\;{\delta(\omega+E_{N}-H)\over
k^{0}-\omega+i\eta\epsilon(k^{0})}\eqno(15)$$
In the same way the Fourier transform of the second term in (10) is
$$\int_{-\infty}^{\infty}dt\;e^{ik^{0}t}
\langle N|\phi(0)\theta_{N}(-t)\phi(x)|N\rangle
=-i\langle N|\phi(0)\;S\;\phi(\vec{x})|N\rangle$$
where
$$S=\int_{-\infty}^{\infty}d\omega\;{\delta(\omega-E_{N}+H)\over
k^{0}-\omega+i\eta\epsilon(k^{0})}\eqno(16)$$
The propagator (10) thus has the form
$$D_{F}(k^{0},\vec{x})=\int_{-\infty}^{\infty}d\omega\;\sum_{N}{e^{-\beta
E_{N}}
\over Z}
{f_{N}(\omega,\vec{x})\over
k^{0}-\omega+i\eta\epsilon(k^{0})}\eqno(17)$$
where
$$\eqalign{f_{N}(\omega,\vec{x})=&\langle
N|\phi(\vec{x})\delta(\omega+E_{N}-H)\phi(0)|N\rangle\cr
 -&\langle
N|\phi(0)\delta(\omega-E_{N}+H)\phi(\vec{x})|N\rangle\cr}$$
It is easy to see that  $f_{N}$ is  the Fourier transform of the commutator:
$$\eqalign{ f_{N}(\omega,\vec{x})= \int_{-\infty}^{\infty}{dt\over 2\pi}\;
e^{i\omega t} \Bigl(&\langle N|\phi(\vec{x})e^{i(E_{N}-H)t}\phi(0)|N\rangle \cr
-&\langle N|\phi(0)e^{i(H-E_{N})t}\phi(\vec{x})|N\rangle\Bigr)\cr
=\int_{-\infty}^{\infty}{dt\over 2\pi}\; e^{i\omega t}
&\langle N|\bigl[\phi(x),\phi(0)\bigr]|N\rangle\cr}$$
where the time dependence of $\phi(x)$ was used in the last step.
Thus the integrand of (17) contains the thermal spectral function:
$$\sum_{N}{e^{-\beta E_{N}}\over Z}f_{N}(\omega,\vec{x})
=\int_{-\infty}^{\infty}dt e^{i\omega t}\rho(t,\vec{x})
=\rho(\omega,\vec{x})\eqno(18)$$
Consequently
$$D_{F}(k^{0},\vec{x})=\int_{-\infty}^{\infty}d\omega\;
{\rho(\omega,\vec{x})\over
k^{0}-\omega+i\eta\epsilon(k^{0})}\eqno(19)$$
The spatial Fourier transform of this is the defining relation (8) and thus
 proves that the operator form (10) is correct.

 To use the operator form
systematically one could  develop the diagramatic rules for perturbation
theory.
This would require a Wick theorem for the $\theta_{N}$ product and an extension
of
 the operator approach of Nieves [8] from the time-ordered to the Feynman case.
It might also be interesting to find the operator form for the finite
temperature
Feynman propagator of quarks and gluons.

\bigskip

\centerline{ACKNOWLEDGMENTS}

It is a pleasure to thank Rob Pisarski and the theory group at Brookhaven
National
Laboratory, where this work was completed, and St\'{e}phane Peign\'{e} for his
helpful comments. This work was supported in part by National Science
Foundation
grant PHY-9213734.

\vfill\eject
\centerline{APPENDIX A: THE FEYNMAN PROPER SELF-ENERGY}

It is  possible to deduce the operator form for the Feynman proper self-energy
$\Pi_{F}(x)$ that generates  $D_{F}(x)$ .
In momentum space the proper self-energy $\Pi_{F}(k)$ is defined by
$$D_{F}(k)={1\over k^{2}-m^{2}-\Pi_{F}(k)}\eqno(A1)$$
The dispersion relation (9) guarantees that $\Pi_{F}(k)$ is analytic in
the region Re$(k^{0})^{2}>0$ and that
 ${\rm Im}\Pi_{F}(k)$ is negative when $k^{0}$ real. Rewrite (A1)
in the form
$$(-k^{2}+m^{2})D_{F}(k)=-1-\Pi_{F}(k)\;D_{F}(k)\eqno(A2)$$
The coordinate space  form of (A2) is
 the Schwinger-Dyson equation:
$$(\square +m^{2})D_{F}(x)=-\delta^{4}(x)-\int
d^{4}y\;\Pi_{F}(x-y)D_{F}(y)\eqno(A3)$$
To deduce the operator form for $\Pi_{F}(x)$ we therefore need to apply
$(\square +m^{2})$ to the operator representaion (10).
 A single time derivative of (10) gives
$$\eqalign{i\dot{D}_{F}(x)=\sum_{N}\Bigl[&\langle
N|\dot\phi(x)\;\theta_{N}(t)\;\phi(0)|N\rangle\cr
 +&\langle
N|\phi(0)\;\theta_{N}(-t)\;\dot\phi(x)|N\rangle\Bigr]
\;{e^{-\beta E_{N}}\over Z}\cr}\eqno(A4)$$
Because of (13) the terms involving the time derivative of
$\theta_{N}(t)$ give
$$\delta(t)\sum_{N}\langle N|\bigl[\phi(x),\;\phi(0)\bigr]|N\rangle
\;{e^{-\beta E_{N}}\over Z}=0\eqno(A5)$$
which contains the commutator is at space-like separation and thus
vanishes by causality. The second time derivative  is
$$\eqalign{i\ddot{D}_{F}(x)=\sum_{N}\Bigl[ \delta(t)&\langle
N|\bigl[\dot\phi(x),
\phi(0)\bigr]|N\rangle\cr +&\langle
N|\ddot\phi(x)\;\theta_{N}(t)\;\phi(0)|N\rangle\cr
 +&\langle
N|\phi(0)\;\theta_{N}(-t)\;\ddot\phi(x)|N\rangle\Bigr]
\;{e^{-\beta E_{N}}\over Z}\cr}\eqno(A6)$$
The first term of (A6) gives $-i\delta^{4}(x)$ because of the equal time
commutation relations. The remaing two terms are given by the field equation
$$\ddot\phi(x)=(\nabla^{2}-m^{2})\phi(x)+J(x)\eqno(A7)$$
where $J=\delta{\cal L}_{I}/\delta \phi$.
Consequently (A6) becomes
$$\eqalign{(\square +m^{2})D_{F}(x)=-\delta^{4}(x)
-i\sum_{N}\Bigl[&\langle N|J(x)\;\theta_{N}(t)\;\phi(0)|N\rangle\cr
 +&\langle
N|\phi(0)\;\theta_{N}(-t)\;J(x)|N\rangle\Bigr]
\;{e^{-\beta E_{N}}\over Z}\cr}\eqno(A8)$$
This is just the form (A3). If $D_{F}$ in (A3) were written as a function of
$x-z$ instead of $x$ only, then the  integrand of the $d^{4}y$ integration
would
be $\Pi_{F}(x-y)D_{F}(y-z)$. Comparison with (A8) gives
 $$\eqalign{\int d^{4}y\,\Pi_{F}(x-y)D_{F}(y-z)=
i\sum_{N}\Bigl[&\langle
N|J(x)\;\theta_{N}(x^{0}-z^{0})\;\phi(z)|N\rangle\cr
 +&\langle
N|\phi(z)\;\theta_{N}(z^{0}-x^{0})\;J(x)|N\rangle\Bigr]
\;{e^{-\beta E_{N}}\over Z}\cr}\eqno(A9)$$
This defines the operator form of the Feynman proper self-energy in direct
analogy to the definition of the time-ordered proper self-energy [10].

\vfill\eject

\centerline{REFERENCES}

\item{1.} L. Dolan and R. Jackiw, Phys. Rev. D9, 3320 (1974).

\item{2.} N.P. Landsman and Ch.G. van Weert, Phys. Rep. 145C, 141 (1987).

\item{3.} H. Umezawa, H. Matsumoto, and M. Tachiki, {\it Thermo Field Dynamics
and Condensed States} (North-Holland, Amsterdam, 1982).

\item{4.} A. J. Niemi and G.W. Semenoff, Ann. Phys. (NY) 152, 105 (1984);
Nucl. Phys. B230, 181 (1984).

\item{5.} R. L. Kobes and G.W. Semenoff, Nucl. Phys, B260, 714 (1985).

\item{7.} H. Matsumoto, I. Ojima, and H. Umezawa, Ann. Phys. (NY) 152, 348
(1984).

\item{8.} J.F. Nieves, Phys. Rev. D42, 4123 (1990).

\item{9.} P. Aurenche and T. Becherrawy, Nucl. Phys. B379, 259 (1992).

\item{10.} S. Mr\'{o}wczy\'{n}ski and U. Heinz, Ann. Phys. (NY) 229, 1 (1994).

\end